\documentclass[aps,prd,groupedaddress,showpacs,graphicx,nofootinbib]{revtex4}
\usepackage{amssymb,graphics,graphicx,color}

\begin{document}

\title{Passive Curvaton}

\author{Chia-Min Lin}\email{cmlin@boar.kobe-u.ac.jp}

\affiliation{Department of Physics, Kobe University, Kobe 657-8501, Japan}

\begin{abstract}
We propose a class of curvaton models which we call passive curvaton. In this paper, two kinds of passive curvaton is considered. The first one is a pseudoscalar curvaton couples to a gauge field. Different from the inflaton case, the constraint from formation of primordial black holes (PBHs) is much weaker and large non-gaussianity (of the equiliteral type) can be produced. The second model is a dilaton-like scalar curvaton couples to a gauge field. We investigate the scale dependence of non-gaussianity in this model. In both models, the spectrum and non-Gaussianity are enhanced by the slow-roll parameter of the curvaton field.
Other possible passive curvaton models are also mentioned.
\end{abstract}
\maketitle

\section{Introduction}
Inflation has became very popular as a theory of the very early universe (for a textbook, see \cite{Lyth:2009zz} and references therein). Unfortunately we still do not really know what causes inflation or what is the inflaton if inflation is driven by a scalar field. For single field inflation case, the universe is well discribed by the 'separate universe' approach \cite{Wands:2000dp}. Namely each Hubble patch can be regarded as a small Friedmann-Roberson-Walker (FRW) universe and each of them 'looks the same' (adiabatic condition) in spite of a possible 'time shift' of its history\footnote{This is an intuitive way of explanation but of course we actually has to be careful about the issue of gauge in general relativity.}. The 'difference' of the number of e-folds of their evolution would corresponds to the curvature perturbation (so-called $\delta N$ formalism \cite{Sasaki:1995aw, Lyth:2005fi}, see \cite{Sugiyama:2012tj} for a recent review). In this picture curvature perturbation is a constant (to any order of perturbation theory) on superhorizon scales due to the identity of each of the 'small universe'. Indeed, if each small universe evolves in the same way, intuitively an extra number of e-folds is not going to change with time. However, this picture can be changed if there exit another field, say a curvaton. Intuitively the independent fluctuation of the curvaton field will give each of the small universe an additional (and independent) contribution of energy density therefore they are not identical anymore (non-adiabatic) and the super-horizon curvature perturbation will evolve with time. If the condition is properly chosen, the curvature perturbation will increase from practically zero to the value we need to explain CMB temperature fluctuation and primordial density perturbation. This is what happens in the case of the curvaton scenario \cite{Lyth:2001nq, Enqvist:2001zp, Moroi:2001ct}. Therefore initially the fluctuation of the curvaton corresponds to isocurvature perturbation (which means no curvature perturbation) at the beggining and finally the isocurvature perturbation is transformed to curvature perturbation. 

For single-field slow-roll inflation with inflaton field $\phi$, the spectrum is given by
\begin{equation}
P^{1/2}=\zeta=\frac{H}{\dot{\phi}}\delta \phi =\frac{1}{2\sqrt{6} \pi M_P^2}\sqrt{\frac{V}{\epsilon}},
\label{inflaton}
\end{equation}
where the slow-roll parameter is 
\begin{equation}
\epsilon \equiv \frac{M_P^2}{2}\left( \frac{V_{,\phi}}{V} \right)^2,
\end{equation}
and $M_P=2.4 \times 10^{18}$ GeV is the reduced Planck scale.


On the other hand, for a curvaton field $\sigma$, 
\begin{equation}
P^{1/2} \equiv r \frac{\delta \sigma}{ \sigma}= r \eta_\sigma \frac{H}{\dot{\sigma}}\delta \sigma,
\label{curvaton}
\end{equation}
where $\sigma$ is the field value of curvaton at horizon exit and $0.01 \lesssim r \lesssim 1$ is roughly the ratio of the curvaton energy density to the total energy density of the universe when it decays \footnote{In order to simplify the expression of equations, in this paper we define $r \equiv \frac{2}{3}\Omega$, where $\Omega \equiv \frac{3\rho_\sigma}{4\rho_\gamma + 3 \rho_\sigma}$ in which $\rho_\gamma$ and $\rho_\sigma$ respectively denote energy density of radiation $\gamma$ and curvaton at curvaton decay. In many literatures, $r$ is used for $\Omega$.}. The lower bound of $r$ is from the constraint of non-Gaussianity of Cosmic Microwave Background (CMB) from WMAP \cite{Komatsu:2010fb}.
The second equality in Eq.~(\ref{curvaton}) holds for a quadratic potential of $\sigma$ with mass $m$ and $\eta_\sigma \equiv m^2/3H^2$ is a slow-roll parameter of the curvaton. This form will become useful in the following sections.

\section{Passive Curvaton}

Let us consider a curvaton field $\sigma$ couples to
a certain quantum field $\chi$. The Lagrangian that is relevant to
$\sigma$ is given by
\begin{equation}
{\cal L}=\frac{1}{2}g^{\mu\nu}\partial_{\mu}\sigma\,\partial_{\nu}\sigma
             -V(\sigma) + {\cal L}_I(\sigma,\chi),
\end{equation}
where $V$ is the curvaton potential, ${\cal L}_I$ is the interaction term,
and the metric is
\begin{equation}
ds^{2}=dt^2-a^2(t) d{\bf x}^2=a^2(d\tau^2-d{\bf x}^2),
\end{equation}
where $\tau$ is the conformal time.
The equations of motion for the mean fields are then given by
\begin{equation}
\ddot{\sigma}+3H\dot{\sigma}+ V^\prime=\frac{\partial {\cal L}_I}{\partial\sigma},
\label{backreaction}
\end{equation}
where $\rho_\chi$ is the energy density of $\chi$ particles,
the dot and the prime denote differentiating with respect to
$t$ and $\sigma$ respectively. In Eq.~(\ref{backreaction}), the right hand
side of the equation is the back reaction of the interaction to
the inflaton mean field. The back reaction arises due to copious
production of $\chi$ quanta during inflation.

In addition, the fluctuations of $\sigma$ satisfy
\begin{equation}
\ddot{\delta\sigma}+3H\dot{\delta\sigma} -\frac{\nabla^{2}}{a^2}{\delta\sigma}+ V^{\prime\prime}{\delta\sigma}=
\delta\left(\frac{\partial {\cal L}_I}{\partial\sigma}\right).
\label{flucteqn}
\end{equation}
The homogeneous solution of this fluctuation equation
gives rise to the standard primordial density fluctuations. Here we call them
as active fluctuations and denote their power spectrum by $P_a$.
The right hand side of Eq.~(\ref{flucteqn}), which comes from the fluctuations of
the backreaction, acts as a source for generating additional fluctuations of $\phi$.
The particular solution of Eq.~(\ref{flucteqn}) with the source term is referred
as passive fluctuations and the power spectrum is denoted by $P_p$. Hence, from Eq.~(\ref{curvaton}) the total
power spectrum is given by the contributions from both active fluctuations and passive fluctuations as
\begin{equation}
P=P_a+P_p=\left(\frac{r}{\sigma}\right)^2 \langle |\delta\sigma_k|^2 \rangle,
\label{totp}
\end{equation}
where $\delta\sigma_k$ is the Fourier mode of $\delta\sigma$ and $P_a$ is given by
\begin{equation}
P_a=\left(\frac{r}{\sigma}\right)^2 \left( \frac{H}{2\pi}\right)^2.
\label{eq26}
\end{equation}
For comparison, for an inflaton we would have (from Eq.~(\ref{inflaton}))
\begin{equation}
P=P_a+P_p=\left(\frac{H}{\dot{\phi}}\right)^2 \langle |\delta\phi_k|^2 \rangle,
\label{totpi}
\end{equation}
One benefit to consider the curvaton scenario is, different from the inflaton case, in Eq.~(\ref{flucteqn}), we do not need to worry about the effect of scalar metric perturbations because the energy density of the curvaton is subdominant at horizon exit and the fluctuation corresponds to isocurvature perturbation. 

\section{Pseudoscalar Curvaton}

For our first example, we consider the case in which the curvaton $\sigma$ is a pseudo Nambu-Goldstone boson (PNGB) with a typical potential after the shift symmetry is broken,
\begin{equation}
V(\sigma)=\Lambda^4 [1-\cos(\sigma/f)],
\label{axion}
\end{equation}
where $\Lambda$ is a mass scale and $f$ is the axion decay constant. We consider a coupling to a gauge field given by
\begin{equation}
{\cal L}_I = -\frac{\alpha}{4f}\sigma F^{\mu\nu}\widetilde{F}_{\mu\nu},
\label{int01}
\end{equation}
where $F_{\mu\nu}=\partial_\mu A_\nu -\partial_\nu A_\mu$ is the field strength associated to some $U(1)$ gauge field $A_\mu$ and $\widetilde{F}^{\mu\nu}$ is its dual.
For simplicity, we approximate Eq.~(\ref{axion}) by a quadratic potential $V=m^2 \phi^2/2$ with $m=\Lambda^2 /f$ which is a good approximation when $\phi \ll f$.

In order to calculate the spectrum, we substitute Eq.~(\ref{int01}) into Eq.~(\ref{flucteqn}) and obtain

\begin{equation}
\left[ \frac{\partial^2}{\partial\tau^2}+2{\cal H} \frac{\partial }{\partial \tau} - \nabla^2   +a^2m^2
\right] \delta \sigma(\tau,{\bf x} )=a^2\frac{\alpha}{f}\left( \vec{E}\cdot\vec{B} - \langle \vec{E}\cdot\vec{B} \rangle \right),
\label{eqx3}
\end{equation}
where we have used conformal time $\tau$ and ${\cal H} \equiv da/(a d\tau)$.
The homogeneous solution corresponds to active fluctuation and the particular solution corresponds to passive fluctuation.
Working in Coulomb gauge, the circular polarization modes obeying \cite{Anber:2009ua}
\begin{equation}
\left[ \frac{\partial^2}{\partial \tau^2}+k^2 \pm \frac{2k\xi}{\tau} \right]A_{\pm}(\tau, k)=0,
\end{equation}
where
\begin{equation}
\xi \equiv \frac{\dot{\sigma}\alpha}{2Hf},
\end{equation}
and dot denotes differentiation with respect to $t$. There is a growth solution of fluctuations described by \cite{Anber:2009ua}
\begin{equation}
A_{+}(\tau,k)~\frac{1}{\sqrt{2k}}\left( \frac{k}{2\xi aH} \right)^{1/4}e^{\pi \xi -2 \sqrt{2\xi k/(aH)}}.
\label{eq36}
\end{equation}
This expression is valid for $\xi \gtrsim \mathcal{O} (1)$.

For the inflaton case\footnote{In this paper, the inflaton case refers to the case where $\chi$ couples to the inflaton instead of curvaton.}, equation similar to Eq.~(\ref{eqx3}) has been solved \cite{Barnaby:2010vf, Barnaby:2011vw} \footnote{ The calculation is rather lenthy and there is no point to repeat it here. We refer the reader to the original references.} by using Eq.~(\ref{eq36}) and the spectrum is (approximately) given by\footnote{We use the same notation for the inflaton case and curvaton case for comparison, but it should be clear through the context concerning which one we are refering to.} 
\begin{equation}
P=P_a \left( 1+ 7.5 \times 10^{-5}\frac{H^2}{4 \pi^2}\left( \frac{\alpha}{2 f} \right)^2 \frac{1}{\xi^2}\frac{e^{4\pi \xi}}{\xi^6} \right),
\label{inflatontot}
\end{equation}
where 
\begin{equation}
P_a=\left( \frac{H^2}{2 \pi \dot{\phi}} \right)^2
\label{inflatona}
\end{equation}
is the active spectrum and
\begin{equation}
\xi \equiv \frac{\dot{\phi}\alpha}{2Hf}=\sqrt{\frac{\epsilon}{2}}\frac{\alpha M_P}{f}.
\label{inflatonxi}
\end{equation}
We can relate Eqs.~(\ref{inflatona}) and (\ref{inflatonxi}) via
\begin{equation}
P_a=\frac{H^2}{4 \pi^2}\left( \frac{\alpha}{2 f} \right)^2 \frac{1}{\xi^2}
\label{eqx}
\end{equation}
and write Eq.~(\ref{inflatontot}) as
\begin{equation}
P=P_a \left( 1+ 7.5 \times 10^{-5}P_a\frac{e^{4\pi \xi}}{\xi^6} \right),
\label{inflatontot2}
\end{equation}

However, this result cannot be applied directly to curvaton. The reason is that the second equality of Eq.~(\ref{inflatonxi}) no longer holds true because curvaton does not drive inflation. This fact will make a crucial difference between curvaton and inflaton case. 

The active spectrum of curvaton is given by Eqs.~(\ref{curvaton}) and (\ref{eq26}) as
\begin{equation}
P_a=r^2\eta^2 \frac{H^4}{4\pi^2 \dot{\sigma}^2}.
\label{cur}
\end{equation}

The relation between curvaton active spectrum and $\xi$ is
\begin{equation}
\frac{P_a}{r^2\eta^2_\sigma}=\frac{H^2}{4 \pi^2}\left( \frac{\alpha}{2 f} \right)^2 \frac{1}{\xi^2}.
\label{eqx2}
\end{equation}
Note that this form is different from Eq.~(\ref{eqx}).
The spectrum of the curvaton is given by substituting Eq.~(\ref{eqx2}) into Eq.~(\ref{inflatontot})
with $P_a$ is given by Eq.~(\ref{cur}), 
\begin{equation}
P=P_a \left( 1+ \frac{7.5 \times 10^{-5}}{r^2 \eta_\sigma^2}P_a\frac{e^{4\pi \xi}}{\xi^6} \right).
\label{pseudo}
\end{equation}
The reason Eq.~(\ref{inflatontot}) can be used here for the curvaton (with different $P_a$) is because the field fluctuation satisfy the same equation. The difference is the way to 'convert' the spectrum of field fluctuation to the spectrum of curvature perturbation is different between inflaton and curvaton cases (see Eqs.~(\ref{totp}) and (\ref{totpi})).


The three point function $\langle \zeta^3 \rangle$ is proportional to the cube of the particular solution of Eq.~(\ref{eqx3}) which is proportioanl to $(\alpha/f)$ from the source term on the right hand side, namely we would have $\langle \zeta^3 \rangle \propto (\alpha/f)^3$. More precisely, it is given by \cite{Barnaby:2011vw}
\begin{eqnarray}
\langle \zeta^3 \rangle &\sim& \frac{3}{10}(2 \pi)^{5/2}P_a^{3/2} \frac{1}{\xi^3} \frac{H^3}{8 \pi^3} \left( \frac{\alpha}{2f} \right)^3  \frac{2.8 \times 10^{-7}}{\xi^9}e^{6 \pi \xi} \label{eq25}  \\
&\equiv& \frac{3}{10}(2 \pi)^{5/2} f_{NL} P^2 
\label{eq26}.
\end{eqnarray}
In this case the curvature perturbations arises from the three point function of curvaton field fluctuation at horizon crossing, therefore the shape of non-Gaussianity is equilateral.
By using Eq.~(\ref{eqx2}) to write $(\alpha/f)$ in terms of $P_a$, we expect to have $\langle \zeta^3 \rangle \propto (1/r \eta_\sigma)^3$. From Eqs.~(\ref{eq25}) and (\ref{eq26}), after
imposing CMB normalization $P=5 \times 10^{-5}$ \cite{Komatsu:2010fb} we obtain
\begin{equation}
f_{NL}^{equil} \sim 4.4 \times 10^{10}\frac{P_a^3}{r^3\eta_\sigma^3}\frac{e^{6 \pi \xi}}{\xi^9}.
\end{equation}

As we can see from Eq.~(\ref{pseudo}), the passive fluctuation (corresponds to the second term) in the curvaton case is enhanced by the slow-roll parameter $\eta_\sigma$ and $r$. However, in general $\xi$ is expected to be smaller than the inflaton case because the Hubble parameter is determined by the inflaton potential instead of the curvaton potential therefore the enhancement is not necessarily large. But it does implies the effect of passive fluctuation can still be significant with a relatively smaller $\xi$ compared with the inflaton case.

As pointed out by \cite{Lin:2012gs}, for the inflaton case, primordial black holes may form. It can be seen from Eq.~(\ref{inflatonxi}) that
 $\xi$ which appears in the exponent of passive fluctuation is proportional to $\sqrt{\epsilon}$. However, for the active fluctuation of inflaton, as shown in Eq.~(\ref{inflaton}), it is proportional to the inverse of  $\sqrt{\epsilon}$. In order to have slow-roll inflation at all, we need $\epsilon \ll 1$ during inflation. But inflation must come to an end therefore $\epsilon$ is expected to increase toward the end of inflation. This implies although it is difficult to have PBH formation from active fluctuation, it would be possible for passive fluctuation.
The constraints from PBH formation could be stronger than non-gaussianity which means non-gaussianity may be negligible. 

However, this is not the case for the curvaton becasue there is no direct relation between $\xi$ and $\epsilon$ like in Eq.~(\ref{inflatonxi}) 
and unless $\dot{\sigma}$ increases dramatically near the end of inflation which is not expected to happen, PBH will not form and large equiliteral shape non-Gaussianity can appear without constraint from PBH formation. 
\section{Scalar Curvaton}
For a second example of passive curvaton, we consider 
\begin{equation}
{\cal L}_I= -\frac{I^2(\sigma)}{4}F^{\mu\nu}F_{\mu\nu},
\end{equation}
where $I(\sigma)$ plays the role of a field dependent gauge coupling.
Now the equation we need to solve is 
\begin{equation}
\left[ \frac{\partial^2}{\partial\tau^2}+2{\cal H} \frac{\partial }{\partial \tau} - \nabla^2   +a^2m^2
\right] \delta \sigma(\tau,{\bf x} )=a^2\frac{I_{,\sigma}}{I}[\vec{E}^2-\vec{B}^2].
\label{eqx4}
\end{equation}
This is more difficult to solve than Eq.~(\ref{eqx3}) because the source term contains function of $\sigma$. The inflaton case is considered in \cite{Barnaby:2012tk}. However, because the inflaton dominates the energy density of the universe during horizon exit, the fluctuation of the inflaton causes fluctuation of the metric tensor therefore Eq.~(\ref{eqx4}) have to be replaced by a more complicated version to take care of the metric perturbation.  

For the inflaton case, in order to simplify the calculation, we can choose the inflaton potential $V(\phi)=M^2\phi^2/2$ and $I \propto \exp(-\phi^2)$ to make $I \propto a^2$. For the curvaton case, we may assume both the inflaton potential and curvaton potential are quadratic, therefore we have $3H\dot{\phi}=-M\phi$ and $3H\dot{\sigma}=-m\sigma$ and by deviding two equations to eliminate $t$ we can show $\phi \propto \sigma$. Consequently we might also choose  $I \propto \exp(-\sigma^2)$ to make $I \propto a^2$. This simplification also allows us to solve the vector modes. This exponential form (dilaton-like interaction) of $I$ is typical in string theory and supergravity.
For the inflaton case we have
\begin{equation}
\frac{I_{,\phi}}{I}=\frac{\dot{I}}{I \dot{\phi}} = 2\frac{H}{\dot{\phi}}= \frac{4\pi M_P}{H}P_a^{1/2}=  \sqrt{\frac{2}{\epsilon}},
\label{eq30}
\end{equation}
where we have used $I \propto a^2$.
Once again the last equality does not hold for the curvaton and we have
\begin{equation}
\frac{I_{,\sigma}}{I}=\frac{\dot{I}}{I \dot{\sigma}} = 2\frac{H}{\dot{\sigma}}= \frac{4\pi M_P}{H}\frac{P_a^{1/2}}{r\eta_\sigma}.
\label{eq31}
\end{equation}
From Eq.~(\ref{eqx4}), we can see these difference makes the source term and the particular solution which corresponds to passive fluctuation different.

The curvature perturbation in the inflaton case is given by \cite{Barnaby:2012tk}
\begin{equation}
P=P_a \left[ 1+192 P_a N^2(N_{tot}-N) \right].
\label{scalar}
\end{equation}
The suppression by powers of the factor $P_a \ll 1$ arises because we are considering higher order perturbation. The factors of $N$ arise due to the growth of inflaton field fluctuation outside the horizon sourced by the gauge field fluctuation and the factor $N_{tot}-N$ is related to the phase space of gauge field fluctuation. Note that this superhorizon growth does not happen to pseudoscalar curvaton (or inflaton) due to different equations of motion. 

For curvaton, the two point function of passive fluctuation is proportional to the square of the source term on the right hand side of Eq.~(\ref{eqx4}). Therefore we have an extra factor of $(1/r\eta_\sigma)^2$  for the curvature perturbation according to Eq.~(\ref{eq31}), namely
\begin{equation}
P=P_a \left[ 1+\frac{192}{r^2\eta^2_\sigma} P_a N^2(N_{tot}-N) \right].
\label{scalarc}
\end{equation}

Similarly, the three point function is proportional to the cube of the source term on the right hand side of Eq.~(\ref{eqx4}). Therefore for the inflaton case, we expect $\langle \zeta^3 \rangle \propto (\sqrt{2/\epsilon})^3$ as can be seen from Eq.~(\ref{eq30}). More precisely we have
\begin{equation}
\langle \zeta^3 \rangle \sim (2\pi)^3P_a^{3/2}\times \frac{27}{8 \pi^{9/2}}H^6\left( \sqrt{\frac{2}{\epsilon}} \right)^3  N^3(N_{tot}-N)  =  (2\pi)^3P_a^3 144 \sqrt{\frac{2}{\pi}}  N^3(N_{tot}-N).
\end{equation} 
The factor $N^3$ is due to the super-horizon growth of the three modes used in computing the three point function. The factor $N_{tot}-N$ is again due to the phase space of gauge field fluctuation.
For curvaton the source term is given by Eq.~(\ref{eq31}) and we have
\begin{equation}
\langle \zeta^3 \rangle \sim (2\pi)^3\left( \frac{P_a}{r\eta_\sigma} \right)^3 144 \sqrt{\frac{2}{\pi}} N^3(N_{tot}-N).
\label{eq35}
\end{equation}
In this case, due to the superhorizon growth of the (nonlinear) curvaton field fluctuation, the non-Gaussianity is close to a local shape.
Compare this with the definition of $f_{NL}$ as in Eq.~(\ref{eq26}),
we obtain \footnote{To be precise, here the non-Gaussianity is not exactly of a local shape and $f_{NL}^{equiv. local}$ is the ``equivalent" local nonlinearity parameter. It is defined so that the average is equal to the local shape non-Gaussianity. We have to use $f_{NL}=(3/4)f_{NL}^{equiv.local}$ in order to obtain the correct numerical factor \cite{Barnaby:2012tk}.}
\begin{equation}
f_{NL}^{equiv. local}= \frac{0.7}{r^3\eta_\sigma^3} \left( \frac{N}{60} \right)^3(N_{tot}-N),
\label{ng}
\end{equation}
where we have used CMB normalization $P \sim P_a =(5 \times 10^{-5})^2$.

\begin{figure}[t]
  \centering
\includegraphics[width=0.4\textwidth]{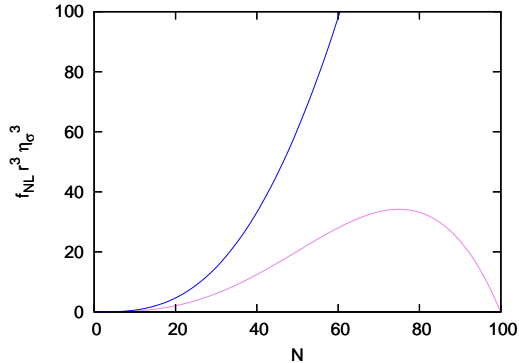}
  \caption{Non-Gaussianity $f_{NL}$ corresponds to number of e-folds $N$. The blue (upper) line corresponds to $N_{tot}=200$ and purple (lower) line corresponds to $N_{tot}=100$. }
  \label{fig2}
\end{figure}

Again for the curvaton case, the passive fluctuation is enhanced by $\eta_\sigma$ and $r$. However,
for $N \sim 60$, the requirement that the first (standard) term in Eq.~(\ref{scalarc}) dominates implies that about $N_{tot}<600$ for the inflaton case and $N_{tot}<600r^2\eta^2_\sigma$ for the curvaton case. The constraint of total number of e-folds becomes more severe for the curvaton case. We plot the result in Fig.~\ref{fig2}. 
As we can see from the plot, the current constaint of non-Gaussianity $f_{NL} \lesssim 100$ from WMAP \cite{Komatsu:2010fb} requires $N_{tot}<200$ for $r \eta_\sigma \sim 1$ and getting more severely for smaller $r \eta_\sigma$. The plot also suggests possibility to have a large running of $f_{NL}$.

However, it can be easily checked that in this case PBHs is not generated near the end of inflation. Indeed when $N \sim 0$ the second term in Eq.~(\ref{scalarc}) vanishes.

For the case of scalar curvaton, although we do not have PBHs formation after inflation, as the inflaton case in \cite{Barnaby:2012tk}, large running of non-Gaussianity may be possible. We will investige this issue in more detail with recent observational constraints. 

If we allow the scale dependence of non-Gaussianity, $f_{NL}$ can be parameterised by using the running index $n$ as

\begin{equation}
f_{NL}(k) \propto k^{n},
\end{equation}
or 
\begin{equation}
n=\frac{d\ln f_{NL}}{d\ln k}=\frac{1}{f_{NL}}\frac{d f_{NL}}{d N}.
\label{run}
\end{equation}

\begin{figure}[t]
  \centering
\includegraphics[width=0.4\textwidth]{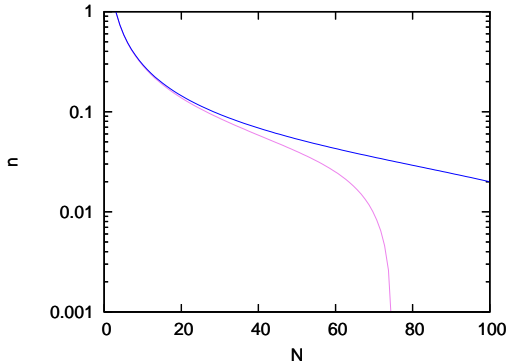}
  \caption{Running non-Gaussianity at horizon exit corresponds to number of e-folds $N$. The blue (upper) line corresponds to $N_{tot}=200$ and purple (lower) line corresponds to $N_{tot}=100$.}
  \label{fig1}
\end{figure}

By using Eqs.~(\ref{ng}) and (\ref{run}), we can calculate $n$ and the result is shown in Fig.~\ref{fig1}.
For $N \sim 60$ the result of running is quite predictive and does not depend on $r \eta_\sigma$. We obtain roughly $0.01 \lesssim n \lesssim 0.1$ and $30 \lesssim f_{NL}  \lesssim 100$ for $100  \lesssim N_{tot}  \lesssim 200$. Although WMAP team assumes a scale independent $f_{NL}$, it is possible to analyse the data assuming varying $f_{NL}$. It is shown in \cite{Becker:2012je} that WMAP data allows running non-Gaussianity as large as $n \sim 0.1$\footnote{To be more accurate, $n=0.3^{+1.9}_{-1.2}$ at $95\%$ confidence.} with $f_{NL} \sim 50$ which is quite close to the result obtained here. The constraint is expected to be improved in near future for example by PLANCK satellite.

\section{Other Possibilities}

The idea of passive curvaton is not restricted to curvaton couples to gauge field. We can consider for example \cite{Barnaby:2009mc, Kofman:2004yc, Green:2009ds, Lee:2011fj},
\begin{equation}
{\cal L}_I=g^2 \sum_i(\sigma-\sigma_i)^2\chi^2_i,
\end{equation}
where $g$ is a coupling constant and $\sigma_i$ is a constant field value. When $\sigma$ rolls down to each trapping point at $\sigma_i$, the $\chi_i$ particles become instantaneously massless and are produced with a number density that increases with $\sigma$'s velocity. As $\sigma$ dumps its kinetic energy into the $\chi_i$ particles, it is slowed down and the produced $\chi_i$ particles are diluted due to the inflationary expansion. However, due to the non-trivial time evolution of $\sigma$, we cannot apply Eq.~(\ref{curvaton}) directly. It would be interesting to consider the detail of this model, but we will not persue it in the current work. 

Another possible direction is to consider models which involve only gravitational interaction without a direct non-gravitational ${\cal L}_I$ between curvaton and other quantum fields. For example we can consider the effect of quantum stress tensor fluctuations \cite{Wu:2006ew, Ford:2010wd}. 

\section{Conclusion and Discussion}
\label{c}

In this paper we consider two models of passive curvaton. The first one is a pseudoscalar curvaton and the second one is a dilaton-like scalar curvaton. We compare the models to the cases where the gauge field couples to an inflaton. We found in general the passive fluctuation is enhanced by the slow-roll parameter of the curvaton. In the first case, PBHs is not expected to form after inflation and non-Gaussianity can well be large. In the second case, we analyse the running of non-Gaussianity.

Passive curvaton in principle can also be applied to the case of inflating curvaton \cite{Dimopoulos:2011gb} (see also \cite{Dimopoulos:2012nj, Furuuchi:2011wa}.) The original inflating curvaton scenario predicts negligible non-Gaussianity because the curvaton dominates the energy density of the universe soon after curvaton start to inflate. However, the idea of passive curvaton can generate non-Gaissianity for inflating curvaton.

For simplicity, we only considered curvaton field fluctuation to first order. However,
in principle if we consider the fluctuation of curvaton field to second order of field fluctuation, Eq.~(\ref{curvaton}) would become
\begin{equation}
\zeta=r \left( \frac{\delta \sigma}{\sigma} + \frac{3}{5}f_{NL}r\left( \frac{\delta \sigma}{ \sigma} \right)^2 \right),
\end{equation}
where
\begin{equation}
f_{NL}=\frac{5}{6r}-\frac{5}{3}-\frac{5}{4}r,
\end{equation}
where $f_{NL}$ represents contribution of higher order fluctuation of $\delta \sigma$ which contains both active and passive fluctuations. Because passive fluctuation is subdominant, $f_{NL}$ roughly corresponds to the contribution of local shape non-Gaussianity from active fluctuation. In addition to that, there will also be cross term between active and passive fluctuations, and higher order term from passive fluctuations.
This can result in non-Gaussianity of a complicated shape which might provide an interesting signal to distinguish our model from others, however the effect will be subdominant and unlikely to be observed in the near future. 

\section*{Acknowledgement}
CML would like to thank Kin-Wang Ng and Kazuyuki Furuuchi for useful discussion.


\newpage
\begin{thebibliography}{99}

\bibitem{Lyth:2009zz}
  D.~H.~Lyth, A.~R.~Liddle,
  Cambridge, UK: Cambridge Univ. Pr. (2009) 497 p.

\bibitem{Wands:2000dp} 
  D.~Wands, K.~A.~Malik, D.~H.~Lyth and A.~R.~Liddle,
  Phys.\ Rev.\ D {\bf 62}, 043527 (2000)
  [astro-ph/0003278].

\bibitem{Sasaki:1995aw} 
  M.~Sasaki and E.~D.~Stewart,
  Prog.\ Theor.\ Phys.\  {\bf 95}, 71 (1996)
  [astro-ph/9507001].

\bibitem{Lyth:2005fi} 
  D.~H.~Lyth and Y.~Rodriguez,
  Phys.\ Rev.\ Lett.\  {\bf 95}, 121302 (2005)
  [astro-ph/0504045].

\bibitem{Sugiyama:2012tj} 
  N.~S.~Sugiyama, E.~Komatsu and T.~Futamase,
  arXiv:1208.1073 [gr-qc].


\bibitem{Lyth:2001nq} 
  D.~H.~Lyth and D.~Wands,
  Phys.\ Lett.\ B {\bf 524}, 5 (2002)
  [hep-ph/0110002].

\bibitem{Enqvist:2001zp} 
  K.~Enqvist and M.~S.~Sloth,
  Nucl.\ Phys.\ B {\bf 626}, 395 (2002)
  [hep-ph/0109214].




\bibitem{Moroi:2001ct} 
  T.~Moroi and T.~Takahashi,
  Phys.\ Lett.\ B {\bf 522}, 215 (2001)
  [Erratum-ibid.\ B {\bf 539}, 303 (2002)]
  [hep-ph/0110096].

\bibitem{Komatsu:2010fb} 
  E.~Komatsu {\it et al.}  [WMAP Collaboration],
  Astrophys.\ J.\ Suppl.\  {\bf 192}, 18 (2011)
  [arXiv:1001.4538 [astro-ph.CO]].



 \bibitem{Anber:2009ua} 
  M.~M.~Anber and L.~Sorbo,
  Phys.\ Rev.\ D {\bf 81}, 043534 (2010)
  [arXiv:0908.4089 [hep-th]].

\bibitem{Barnaby:2010vf}
  N.~Barnaby and M.~Peloso,
  Phys.\ Rev.\ Lett.\  {\bf 106}, 181301 (2011)
  [arXiv:1011.1500 [hep-ph]].

\bibitem{Barnaby:2011vw}
  N.~Barnaby, R.~Namba, and M.~Peloso,
  J. Cosmol. Astropart. Phys. 4 (2011) 9
  [arXiv:1102.4333 [astro-ph.CO]].





\bibitem{Lin:2012gs} 
  C.~-M.~Lin and K.~-W.~Ng,
  arXiv:1206.1685 [hep-ph].





\bibitem{Barnaby:2012tk} 
  N.~Barnaby, R.~Namba and M.~Peloso,
  Phys.\ Rev.\ D {\bf 85}, 123523 (2012)
  [arXiv:1202.1469 [astro-ph.CO]].

\bibitem{Becker:2012je} 
  A.~Becker and D.~Huterer,
  Phys.\ Rev.\ Lett.\  {\bf 109}, 121302 (2012)
  [arXiv:1207.5788 [astro-ph.CO]].

\bibitem{Barnaby:2009mc} 
  N.~Barnaby, Z.~Huang, L.~Kofman and D.~Pogosyan,
  Phys.\ Rev.\ D {\bf 80}, 043501 (2009)
  [arXiv:0902.0615 [hep-th]].

\bibitem{Kofman:2004yc} 
  L.~Kofman, A.~D.~Linde, X.~Liu, A.~Maloney, L.~McAllister and E.~Silverstein,
  JHEP {\bf 0405}, 030 (2004)
  [hep-th/0403001].

\bibitem{Green:2009ds} 
  D.~Green, B.~Horn, L.~Senatore and E.~Silverstein,
  Phys.\ Rev.\ D {\bf 80}, 063533 (2009)
  [arXiv:0902.1006 [hep-th]].

\bibitem{Lee:2011fj} 
  W.~Lee, K.~-W.~Ng, I-C.~Wang and C.~-H.~Wu,
  Phys.\ Rev.\ D {\bf 84}, 063527 (2011)
  [arXiv:1101.4493 [hep-th]].

\bibitem{Wu:2006ew} 
  C.~-H.~Wu, K.~-W.~Ng and L.~H.~Ford,
  Phys.\ Rev.\ D {\bf 75}, 103502 (2007)
  [gr-qc/0608002].

\bibitem{Ford:2010wd} 
  L.~H.~Ford, S.~P.~Miao, K.~-W.~Ng, R.~P.~Woodard and C.~-H.~Wu,
  Phys.\ Rev.\ D {\bf 82}, 043501 (2010)
  [arXiv:1005.4530 [gr-qc]].





\bibitem{Dimopoulos:2011gb} 
  K.~Dimopoulos, K.~Kohri, D.~H.~Lyth and T.~Matsuda,
  JCAP {\bf 1203}, 022 (2012)
  [arXiv:1110.2951 [astro-ph.CO]].

\bibitem{Dimopoulos:2012nj} 
  K.~Dimopoulos, K.~Kohri and T.~Matsuda,
  Phys.\ Rev.\ D {\bf 85}, 123541 (2012)
  [arXiv:1201.6037 [hep-ph]].

\bibitem{Furuuchi:2011wa} 
  K.~Furuuchi and C.~-M.~Lin,
  JCAP {\bf 1203}, 024 (2012)
  [arXiv:1111.6411 [hep-ph]].



\end{thebibliography}
\end{document}